# Km4City Ontology Building vs Data Harvesting and Cleaning for Smart-city Services

Pierfrancesco Bellini, Monica Benigni, Riccardo Billero, Paolo Nesi, Nadia Rauch
DISIT Lab, Dep. of Information Engineering, University of Florence, Italy
http://www.disit.dinfo.unifi.it , {pierfrancesco.bellini, riccardo.billero, paolo.nesi, nadia.rauch}@unifi.it

*Abstract*— **Presently, a very large number of public and private data sets are available from local governments. In most cases, they are not semantically interoperable and a huge human effort would be needed to create integrated ontologies and knowledge base for smart city. Smart City ontology is not yet standardized, and a lot of research work is needed to identify models that can easily support the data reconciliation, the management of the complexity, to allow the data reasoning. In this paper, a system for data ingestion and reconciliation of smart cities related aspects as road graph, services available on the roads, traffic sensors etc., is proposed. The system allows managing a big data volume of data coming from a variety of sources considering both static and dynamic data. These data are mapped to a smart-city ontology, called KM4City (Knowledge Model for City), and stored into an RDF-Store where they are available for applications via SPARQL queries to provide new services to the users via specific applications of public administration and enterprises. The paper presents the process adopted to produce the ontology and the big data architecture for the knowledge base feeding on the basis of open and private data, and the mechanisms adopted for the data verification, reconciliation and validation. Some examples about the possible usage of the coherent big data knowledge base produced are also offered and are accessible from the RDF-Store and related services. The article also presented the work performed about reconciliation algorithms and their comparative assessment and selection.**

## I. Introduction

Despite the large work performed by Public Administrations (PAs) on producing open data they are not typically semantically interoperable each other and neither with the many private data available in the city. Open data coming from PA contains typically statistic information about the city (such as data on the population, accidents, flooding, votes, administrations, energy consumption, presences on museums, etc.), location of point of interests, POIs, on the territory (including, museums, tourism attractions, restaurants, shops, hotels, etc.), major GOV services, ambient data, weather status and forecast, changes in traffic rules for maintenance interventions, etc. Moreover, a relevant role is covered in the city by private data coming from mobility and transport such as those created by Intelligent Transportation Systems, ITS, for bus management, and solutions for managing and controlling parking areas, car and bike sharing, car flow in general, good delivering services, accesses on Restricted Traffic Zone, RTZ, etc. Both open and private data may include real time data such as the traffic flow measure, position of vehicles (buses, car/bike sharing, taxi, garbage collectors, delivering services, etc.), railway and train status with respect to the arrival, park areas status, and Bluetooth tracking systems for monitoring movements of cellular phones as people, ambient sensors, and TV cameras streams for security and flow. Both PAs and mobility operators have large difficulties in elaborating and aggregating data to provide new services, even if they could have a strong relevance in improving the citizens' quality of life and services. Therefore, *our cities are not so smart as they could be* by exploiting a semantically interoperable knowledge base on the available data. This condition is also present in highly active cities on open data publication such as Firenze, Italy, that is considered one of the top cities on Open Data in Italy and in Europe.

Therefore, variability, complexity, variety, and size of these data make the data process of ingestion, aggregation, to enable their exploitation a "Big Data" problem as addressed in [2], [3]. The variety and variability of data can be due to the presence of several different formats, and to scarce (or non-existing) interoperability among semantics of the single fields and of the several data sets. In order to reduce the ingestion and integration cost, by optimizing services and exploiting integrated information at the needed quality level, a better interoperability and integration among systems is required [1], [2]. This problem can be partially solved by using specific reconciliation processes to make these data interoperable with other ingested and harvested data. The velocity of data is related to the frequency of data update, and it allows distinguishing static from dynamic data. Static data are rarely updated, such as once per month/year, as opposed to the dynamic data which are updated: from once a day up to every minute or more, to arrive at real time data. When these data models are analyzed and then processed to become semantically interoperable, they can be used to create an integrated knowledge base that can be feed by corresponding data instances (with static, quasi-static and real time data). On the other hand, this approach does not solve the problem since instances can be not interoperable and linked together. For example, a street names coming from two different sources physically identifying the same street may be written in different manner creating a sematic miss-link. These problems have to be solved as well with reconciliation processes.

The above knowledge model construction may lead to successfully create a large semantically interoperable knowledge base that can be used to provide service to third party applications of public administration or enterprises. These applications can exploit the knowledge base making queries.

For example: searching services around a certain GPS point, looking for area in which restaurant are not available, detecting and predicting critical conditions, computing suggestions for service tuning on the basis of statistical data, deducing of causalities. These services can be contextualized and used by different operators in the city such as: public administrations, mobility operators, and commercials. Moreover, specific applications for alerting on forecast and/or critical conditions are going to be produced providing service to the public administrators.

In this paper, the above mentioned process of knowledge base construction is described from: ontology creation to the data ingestion and knowledge base production and validation. It includes processes of data analysis for ontology modeling, data mining, formal verification of inconsistencies and incompleteness to perform data reconciliation and integration. Among the several issues, the most critical aspects are related to the ontology construction that enables deduction and reasoning, and on the verification and validation of the obtained model and knowledge base. The paper is organized as follows. In Section II, the overview of the proposed ontology is presented together with the main problems underlined its construction, and the main macro classes. Section III describes the details associated with each macroclass of proposed km4City (Knowledge Model for City) ontology and the integration with other vocabularies. Section IV reports the general architecture adopted for processing Open Data and the motivations that constrained its definition. In the same section, two services are presented that allow navigating in the knowledge base and can be used by non-data engineers. Section V presents the verification and validation process adopted for the knowledge base, and the results regarding the reconciliation precision and recall by using different kind of algorithms. Conclusions are drawn in Section VI.

## II. km4City Ontology main elements

In order to create a knowledge model for Smart City services, a large number of data sets have been analyzed to see in detail each single data elements of each single data set with the aim of modeling and establishing the needed relationships among elements, thus making a general data set semantically interoperable at model level (e.g., associating the street names with toponimous coding, resolving ambiguities). The result of this deep analysis phase is the *km4City*, a knowledge model for the city and its services. The work performed started from the models of the data sets available in the Florence and Tuscany area, and from the general datasets for similar data available on the several open data portals. In total the whole data sets taken into account have been are more than 800. At regional level, Tuscany Region provided a set of open data into the MIIC (Mobility Integration Information Center of the Tuscany Region), and provide integrated and detailed geographic information reporting each single street in Tuscany (about 137,745), and the locations of a large part of civic numbers, for a total of 1,432,223 (a wider integration could be performed integrating also Google maps and Yellow/white pages). From the MIIC, it is possible to recover information regarding streets, car parks, traffic flow, bus timeline, etc. While from Florence municipality, real time data such as those from the RTZ about car passages, tram lines on the maps, bus stops, bus tickets, statistics on accidents, ordinances and resolutions, numbers of arrivals in the city, number of vehicles per year, etc., can be obtained. From the other open data, points of interest, POI, can be recovered as position and information related to: museums, monuments, theaters, libraries, banks, express couriers, police, firefighters, restaurants, pubs, bars, pharmacies, airports, schools, universities, sports facilities, hospitals, emergency rooms, government offices, hotels and many other categories, including weather forecast by LAMMA consortium. In addition to these data sets, the private data coming from the mobility and transport operators have been collected as well.

The analysis of the above mentioned data sets allowed us to create the km4City integrated ontological model presenting 7 main areas of macroclasses as depicted in Figure 1, and described as follows.

*Administration*: includes classes related to the structuring of the general public administrations, namely PA, and its specifications, Municipality, Province and Region; also includes the class Resolution, which represents the ordinance resolutions issued by each administration that may change the traffic stream.

*Street-guide*: formed by entities as Road, Node, RoadElement, AdministrativeRoad, Milestone, StreetNumber, RoadLink, Junction, Entry, and EntryRule Maneuver, it is used to represent the entire road system of Tuscany, including the permitted maneuvers and the rules of access to the RTZ. The street model is very complex since it may model from single streets to areas, different kinds of crosses and superhighways, etc. In this case, OTN (Ontology for Transport Network) vocabulary has been exploited to model traffic [4] that is more or less a direct encoding of GDF (Geographic Data Files) in OWL.

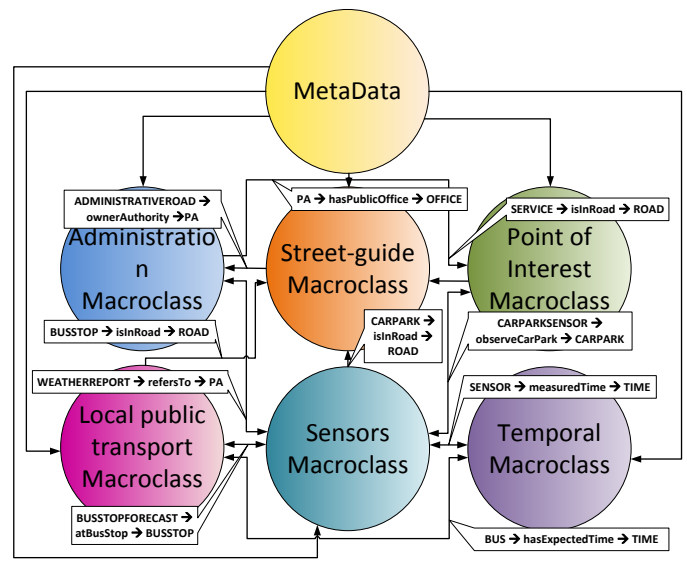

Figure 1 - Ontology Macro-Classes and their connections

*Point of Interest (POI)*: includes all services, activities, which may be useful to the citizen and who may have the need to "search-for" and to "arrive-at". The classification of individual services and activities is based on main and secondary categories planned at regional level. In addition, this macro segment of the ontology may take advantage of reusing Good Relation model of the commercial offers[1]: in fact, the ontological model *km4City* allows connecting Service instances to the corresponding instances of Location belonging to GoodRelations model [16].

*Local public transport*: includes the data related to major LPT (Local Public Transport, in Italian: TPL, Transport Public Local) companies scheduled times, the rail graph, and data relating to real time passage at bus stops. Therefore, this macroclass is formed by classes PublicTransportLine, Ride, Route, AVMRecord, RouteSection, BusStopForeast, Lot, BusStop, RouteLink, RouteJunction. (where AVM means Automatic Vehicle Monitoring).

*Sensors*: macroclass concerns data from sensors: ambient, weather, traffic flow, pollution, etc. Currently, data collected by various sensors installed along some streets of Florence and surrounding areas, and those relating to free places in the main car parks of the region, have been integrated in the ontology. Some of the sensors can be located on moving vehicles such as those on busses, car sharing, bike sharing, and on citizens' mobiles, etc.

*Temporal*: macroclass that puts concepts related to time (time intervals and instants) into the ontology, so that associate a timeline to the events recorded and is possible to make forecasts. It takes advantage from time ontologies such as OWL-Time [5].

*Metadata*: This group of entities represents the collection of metadata associated with the data sets, and their status conditions. If they have been ingested and integrated into the RDF store index, data of ingestion and update, licenses information, versioning, etc. In the case of problems with a certain set of triples or attributes, it is possible to recover the data sets that have generated them, when and how.

The km4City ontology reuses the following vocabularies: *dcterms*: set of properties and classes maintained by the Dublin Core Metadata Initiative; *foaf*: dedicated to the description of the relations between people or groups; *schema.org*: for a description of people and organizations; *wgs84_pos*: vocabulary representing latitude and longitude, with the WGS84 Datum, of geo-objects. The present RDF store and indexing engine OWLIM allows to perform geographic queries, for example to identify the POI which are closer than a given distant with respect to a specific GPS position. To this end, a specific index is built during RDF store indexing.

### III. KM4CITY SMART-CITY ONTOLOGY DETAILS

#### A. Administration Macroclasss

The Administration Macroclass is structured in order to represent the Italian public administration hierarchy: each region is divided into several provinces, within which the territory is divided into municipalities. Moreover each PA, during its mandate, can produce resolutions and publish statistics. To represent this situations the km4City Ontology has, as main class of Administration Macroclass, the class *PA*, which has been defined as a subclass of *foaf:Organization*, link that helps to assign a clear meaning to this class. The three subclasses of *PA*, i.e. *Region*, *Province* and *Municipality* are automatically defined according to the restriction on some ObjectProperties: for example, the class *Region* is defined as a restriction of the class *PA* on ObjectProperty *hasProvince*, so that only the PA that possess provinces, can be classified as *Regions*. Class *PA* is connected to class *Resolution* through the ObjectProperty *hasApprovedPA*, that has its inverse property, *hasResolution*. Statistical data related to both various municipalities in the region and to each street, are represented by a unique class *StatisticalData*, shared by macroclasses Administration and Street Guide: as we will see also in the next subsection, class *StatisticalData* is connected to both classes *PA* and *Road* through ObjectProperty *hasStatistic*.

---
[1] http://www.heppnetz.de/projects/goodrelations/

#### B. Street-guide Macroclass

At regional level, the entire roads system, from an administrative point of view, is seen as a set of administrative extensions or administrative roads, while from the citizen' point of view, it can be regarded as composed by a set of roads. Each *administrative* road represents the administrative division of the roads, based on which the PAs have to manage them. Both administrative roads and roads are formed by a variable number of road elements, each of which starts and ends in a unique node. Each road element, in turn, is formed by a set of sections separated by an initial and a final junction, which allow delineating the exact segmented line representing the road element. The street numbers are placed on the roads, each of which always corresponds to at least one entry. In some cases, there are two entrances which correspond to a single street number, i.e., the outer gate and the front door. With respect to the road circulation, access rules and maneuvers are defined: the first one defines access restrictions to each road element, the seconds are mandatory turning maneuvers, priority or forbidden, which are described by indicating the order of road elements involving.

Another relevant element of the road system is the milestone, which represents the kilometer stones that are placed along the administrative roads. They are elements that identify the precise value of the mileage at that point.

The above described situation has been modeled into the km4City Ontology, choosing as the main class of Street Guide macroclass, the *RoadElement* class, which is defined as a subclass of the corresponding element in the OTN Ontology (see Figure 2), that is *Road_Element*. Each road element is delimited by a start node and an end node, detectable by the ObjectProperties *startsAtNode* and *endsAtNode*, which connect elements of the class in question to the class *Node*, subclass of the same name class *OTN:Node*, belonging to ontology OTN.

The class *Node* has been defined with a restriction on DataProperty *geo:lat* and *geo:long*, two properties inherited from the definition of the class *Node* as subclass of *geo:SpatialThing* belonging to ontology Geo wgs84 [7]: in fact, each node can be associated with only one pair of coordinates in space, and a node without these values cannot exist. The class *Road* is defined as a subclass of the corresponding class in the OTN Ontology, i.e., the homonymous class *Road*, with a cardinality restriction on ObjectProperty *containsElement*, since a road that does not contain at least one road element, cannot exist. Also the class *AdministrativeRoad* is connected to class *RoadElement* through two inverse ObjectProperties *hasRoadElement* and *formAdminRoad*, while it is connected with only one ObjectProperty, *coincideWith*, to the class *Road*.

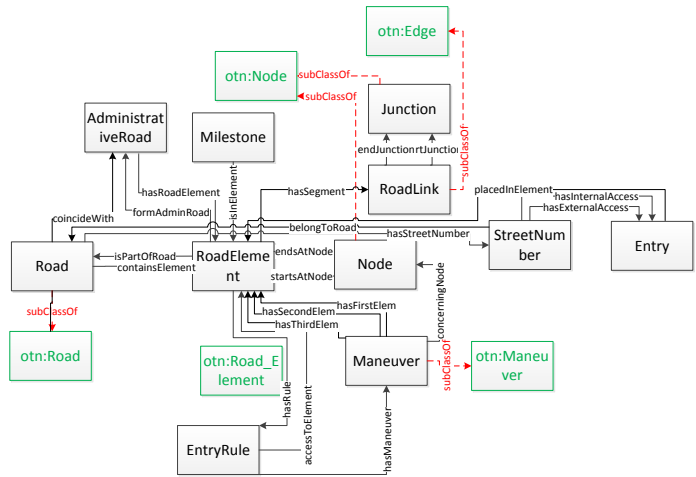

Figure 2 - The Street-guide Macro class of km4City Ontology

In order to better clarify the relationship that exists between classes *Road*, *AdministrativeRoad* and *RoadElement*: a *Road*'s instance can be connected to multiple instances of class *AdministrativeRoad* (e.g., if a road crosses the border between two provinces), but the opposite is also true (e.g., when a road crosses a provincial town center and it assumes different names), i.e., there is a N:M relationship between these two classes. On each road element, it is possible to define access restrictions, identified by class *EntryRule*, which is connected to class *RoadElement* through 2 inverse ObjectProperties, i.e., *hasRule* and *accessToElement*. The class *Maneuver* and class *EntryRule* are connected by ObjectProperty *hasManeuver*. Moreover, we verified that only in rare cases maneuvers involving three different road elements, to represent the relationship between classes *Maneuver* and *RoadElement*, three ObjectProperties were defined: *hasFirstElem*, *hasSecondElem* and *hasThirdElem.* In addition to the ObjectProperty that binds a maneuver to the junction that is interested, that is, *concerningNode* (because a maneuver takes place always in proximity of a node). Each instance of *Milestone* class must be associated with a single instance of *AdministrativeRoad*, and it is therefore defined a cardinality restriction equal to 1. Associated with ObjectProperty *isInElement*; also class *Milestone* is defined as subclass of *geo:SpatialThing*, in this case the presence of coordinates is not mandatory, to be capable to model entities that does not present those data. Thanks to the owned data, classes *StreetNumber* and *Entry* were defined: the connection of class *StreetNumber* to class *Road*, is possible respectively through the ObjectProperties *hasStreetNumber* and *belongToRoad*. The relationship between classes *Entry* and *StreetNumber*, is also defined by the two ObjectProperties, *hasInternalAccess* and *hasExternalAccess*. Class *Entry* is defined as a subclass of *geo:SpatialThing*, and it is possible to associate a maximum of one pair of coordinates *geo:lat* and *geo:long* with each instance. The Street-guide macroclass is connected to the Administration macroclass through two different ObjectProperties -- i.e., *OwnerAuthority* and *managingAuthority*, which represent respectively the public administration which owns an *AdministrativeRoad*, or public administration that manages a *RoadElement*. Thanks to the processing of *KMZ* files (Keyhole Markup Language file and zero or more supporting files packaged in a ZIP file), is possible to retrieve the set of coordinates that define the broken line of each *RoadElement*. Each of these points is added to the ontology as an instance of class *Junction* (defined as a subclass of *geo:SpatialThing*, with compulsory single pair of coordinates). Each small segment between two instances of *Junction* class is instead an instance of class *RoadLink*, which is defined by a restriction on the ObjectProperties *ending* and *starting*, which connect the two mentioned classes. RoadLink and Juctions are in total about 20 million of triples.

### C. Point of Interest Macroclass

This macroclass allows to represent services to the citizens, points of interest, businesses activities, tourist attractions, and anything else can be located thanks to a pair of coordinates on a map. Each type of element has been defined starting from the categories defined by the Tuscany Region taxonomy of categories, including: Accommodation, GovernmentOffice, TourismService, TransferService, CulturalActivity, FinancialService, Shopping, Healthcare, Education, Entertainment, Emergency and WineAndFood. The main class of the POI Macroclass is a generic class *Service* for which the subclasses above listened have been identified thanks to the value assigned to ObjectProperty *serviceCategory*.

The class *Accommodation,* for example, was defined as a restriction of class *Service* on ObjectProperty *serviceCategory*, which must take one of the following values: *tourist_resort*, *hotel*, *tourist_home*, *rest_home*, *religiuos_guest_house*, *bed_and_breakfast*, *hostel*, *summer_residence*, *vacation_resort, farmhouse*, *day_care_center*, *camping*, *historic_residence*, and *mountain_dew*.

We have also defined DataProperty *ATECOcode*, i.e. ATECO is the ISTAT (national institute for statistics in Italy, www.istat.it) code for the classification of economic activities, which could be used in future as a filter to define the various services subclasses, in place of the categories proposed by the Tuscany Region database, in order to make more precise research of the various types of services. Thanks to class *Service,* macroclasses *Point of Interest* and *Street guides* can be connected by exploiting ObjectProperty *hasAccess*, with which a service can be connected to only one external access, corresponding to the road and the street number of the service location. If this association is not possible (because of lack of information, missing street number, etc.), the connection between the same two macroclasses listed above, is realized through the ObjectProperty *isInRoad,* that connects an instance of the class *Service* to an instance of the class *Road*. In order to use at least one of these two ObjectProperty to connect macroclasses *Point of Interest* and *Street Guides*, an intense reconciliation phase is necessary, as described in *section IV*.

As mentioned in the previous paragraph, the *km4City* ontology has the ability to interconnect each instance of the Service class to the corresponding instance defined according to the ontological model GoodRelations, i.e., *gr:Location*; such connection can be finalized using the ObjectProperty *hasGRLocation*.

### D. Public Transport Macroclass

The TPL (Italian LPT) macroclass (see Figure 3) includes information relating to public transport by road and rail. The public transport by road is organized in public transport lots, each of which is composed by a number of bus and tram lines. Each line includes at least two rides per day (the first in ascendant direction, and the second one in descendant direction), identified through a code provided by the TPL company and each ride is scheduled to drive along a specific path, called route. A route can be seen as a series of road segments delimited by subsequent bus stops, but wishing then to represent to a cartographic point of view the path of a bus, we need to represent the broken line that composes each stretch of road crossed by the means of transport itself, and to do so, the previously used modeling on road elements, has been reused: we can see each path as a set of small segments, each of which delimited by two junctions.

The part relating to rail transport: each railway line, i.e., an infrastructure designed to run trains between two places of service, is composed by a number of railway elements, which can also form a railway direction (a railway line having particular characteristics of importance for volume of traffic and transport relations linking centers or main nodes of the rail network) and a railway section (section of the line in which you can find only one train at time, and that is usually preceded by a "protective" or "block" signal). In addition, each rail element begins and ends at a railway junction, in correspondence of which there may be train stations or cargo terminals.

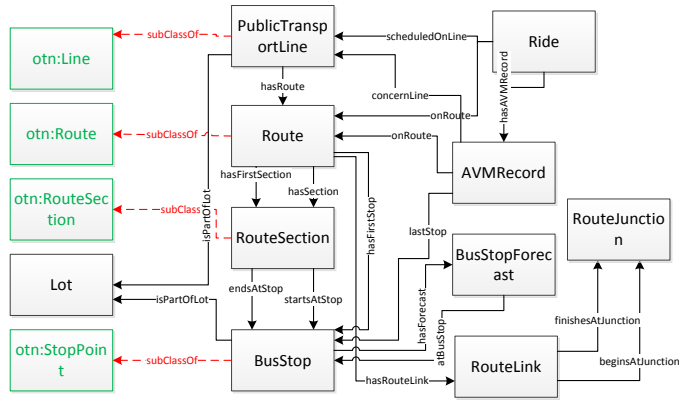

Figure 3 - km4City - Public Transport Macroclass (a portion)

Based on the previous description, we have defined class *PublicTransportLine* (a subclass of *OTN:Line*), which is connected to the corresponding instance of class *Lot*, thanks to ObjectProperty *isPartOfLot*. Every instance of class *PublicTransportLine* is connected to class *Ride* through ObjectProperty *scheduledOnLine*, which is defined as a limitation of cardinality exactly equal to 1, because each stroke may be associated to a single line. To model each path and its sequence of crossed bus stops, classes *Route* and *BusStop* have been defined. It has been decided to define two ObjectProperties linking classes *Route* and *RouteSection*, i.e. *hasFirstSection* and *hasSection*, since, from a cartographic point of view, wanting to represent the path that a certain bus follows. In details, knowing the first segment and the stop of departure, it is possible to obtain all the other segments that make up the complete path and, starting from the second bus stop (that is identified as the different stop from the first stop, but that is also contained in the first segment), we are able to reconstruct the exact sequence of the bus stops, and then the segments, which constitute the entire path. For this purpose, ObjectProperty *hasFirstStop* has been defined, which connects classes *Route* and *BusStop* and ObjectProperty *endsAtStop* and *startsAtStop*, instead of connecting each instance of *RouteSection* to eventual two instances of class *BusStop* (subclass of *OTN:StopPoint*). Each stop is connected to class *Lot*, through the ObjectProperty *isPartOfLot*, with a 1:N relation, because there are stops shared by urban and suburban lines so they belong to two different lots. Possessing also the coordinates of each stop, class *BusStop* was defined as a subclass of *geo:SpatialThing*, and was also termed a cardinality equal to 1 for the two DataProperty *geo:lat* and *geo:long*. In order to represent the broken line that composes each route, classes *RouteLink* and *RouteJunction*, and the ObjectProperties *beginsAtJunction* and *finishesAtJunction*, were defined. Class *Route* is connected to class *RouteLink* through *hasRouteLink* ObjectProperty.

The Railway Graph is mainly formed by class *RailwayElement*, that can be connected to classes *RailwayDirection* and *RailwaySection*, thanks to two inverse ObjectProperties *isComposedBy* and *composeSection*, and to class *RailwayLine*, trough the two inverse ObjectProperties *isPartOfLine* and *hasElement*. Each instance of class *RailwayElement* is connected to two instances of class *RailwayJunction* (defined as a subclass of the OTN:Node), by the ObjectProperties *startAtJunction* and *endAtJunction*,. Classes *TrainStation* and *GoodsYard* correspond only to one instance of the *RailwayJunction* class, both through the ObjectProperty *correspondToJunction*.

### E. Sensors Macroclass

Sensors Macroclass consists of four parts related to car parks sensors, weather sensors, traffic sensors installed along roads/rails and to AVM/kit systems installed on buses, cars and/or bikes. The first part is focused on the real-time data related to parking: for each sensors installed into different car parking areas, a status record is received every 5minutes. In each status report, there is information about the number of free and occupied parking spaces, for the main car parks. The weather sensors produce real-time data concerns the weather forecast, thanks to LAMMA (institute for modeling and monitoring environmental conditions in Tuscany, http://www.lamma.rete.toscana.it). This consortium updates the municipality forecast report once or twice per day and every report contains forecast for five days divided into range, which have a greater precision (and a higher number) for the nearest days until you get to a single daily forecast for the 4th and 5th day. The traffic sensors produce real-time data concerning the sensors placed along the roads of the region, which allow making different measures and assessment related to traffic situation. Unfortunately, the location of these sensors is not very precise, it is not possible to place them in a unique point thanks to coordinate, but only to place them within a toponym, which for long-distance roads such as FI-PI-LI road (the highway that connect Florence-Pisa-Livorno), it represents a range of many miles. Each sensor, is part of a group and produces observations which can belong to four types, i.e. they can be related to the average velocity, car flow passing in front of the sensor, traffic concentration, or to the traffic density. On this regards, Bluetooth sensors could be installed to trace the number of people passing by from a given point.

The AVM (Automatic Vehicle Monitoring) systems part concerns the sensors systems installed on most of buses, which, at intervals of few minutes, send a report to the management center. They provide information about: the last stop performed, current GPS coordinates of the vehicle, the vehicle identifiers and bus line, a list of upcoming stops with the planned passage time.

To model the car parks situation class *CarParkSensor* has been defined which is linked to instances of class *SituationRecord*, that represents, the state of a certain parking at a certain instant. This connection is performed via the reverse ObjectProperties: *relatedToSensor* and *hasRecord*. This first part of the Sensors Macroclass is also connected to the Point of Interest Macroclass through two inverse ObjectProperties: *observeCarPark* and *hasCarParkSensor*, which connect the classes *CarParkSensor* and *TransferService,* respectively.

The weather situation is represented by class *WeatherReport* connected to class *WeatherPrediction* via the ObjectProperty *hasPrediction*. Moreover, class *Municipality* is connected to each report by two reverse ObjectProperties: *refersToMonicipalitu* and *hasWeatherReport*, to realize the connection between the macroclasses Sensors and Administration.

With regard to traffic sensors, each group of sensors is represented by class *SensorSiteTable* and each instance of class *SensorSite* connects to its group through the ObjectProperty *formsTable* and thanks to ObjectProperty *placeOnRoad* each instance of class *SensorSite* can be connected only to class *Road* (see Figure 2), to create a connection between Sensors and Street-guide macroclasses. Each sensor produces observations represented by instance of class *Observation* and, as mentioned earlier, there are four possible subclasses: *TrafficConcentration, TrafficHeadway, TrafficSpeed*, and *TrafficFlow* subclass. Classes *Observation* and *Sensor* are connected via a pair of reverse ObjectProeprties, *hasObservation* and *measuredBySensor*.

Finally, the last part of Sensors Macroclass is mainly represented by two classes, *AVMRecord* and *BusStopForecast*, and thanks to the ObjectProperty *lastStop*, this first class can be connected to the *BusStop* class. The list of scheduled stops is instead represented as instances of the class *BusStopForecast*, a class that is linked to the class *BusStop* through *atBusStop* ObjectProperty so as to be able to recover the list of possible lines provided on a certain stop (the class *AVMRecord* is in fact also connected to the class *Line* via ObjectProperty *concernLine*).

### F. Temporal Macroclass

The Temporal Macroclass, is now only "sketchy" within the ontology, and it is based on the Time ontology [5] as it has been used into OSIM ontology [8]. It requires the integration of the concept of time as it will be of paramount importance to be able to calculate differences between time instants, and the Time ontology comes to help us in this task. We define fictitious URI: *#instantForecast*, *#instantAVM*, *#instantParking*, *#instantWreport*, *#instantObserv* to associate at a resource URI a time parameter -- i.e. respectively *BusStopForecast*, *AVMRecord*, *SituationRecord*, *WheatherReport* and finally *Observation*. It is necessary to create a fictitious URI that links a time instant to each resource, to avoid ambiguity, because identical time instants associated with different resources may be present (although the format in which a time instant is expressed has a fine scale). Time Ontology is used to define precise moments as temporal information, and to use them as extreme for intervals and durations definition, a feature very useful to increase expressiveness.

Pairs of ObjectProperties have also been defined for each class that needs to be connected to the class *Instant*: between classes *Instant* and *SituationRecord* were defined the inverse ObjectProperties *instantParking* and *observationTime*, between classes *WeatherReport* and *Instant*, the ObjectProperties *instantWReport* and *updateTime* have been defined; between classes *Observation* and *Time* there are the reverse ObjectProperties *measuredTime* and *instantObserv*, between *BusStopForecast* and *Time* we can find *hasExpectedTime* and *instantForecast* ObjectProperties, and finally, between *AVMRecord* and *Time*, there are the reverse ObjectProperties *hasLastStopTime* and *instantAVM*.

### G. Metadata Macroclass

Finally, Metadata macroclass is used to keep track of the status and descriptors associated with the various ingested dataset. Sesame [www.openrdf.org] allows assigning a name (i.e., an identifier) to the various graphs that can be identified within the defined ontology, so defining some Named Graphs. This name, also called "context", allows expanding the triple data model to a quad data model, defined as follow: subject-predicate-object-*context*. Owlim, allows to assign the context to each triple set, during the data loading phase. Therefore, a description and status context called *dataProperty* is associated with each data set. It allows to store all the useful information related to a certain data set, such as: date of creation, data source, original file format, dataset description, type of license bound to the dataset, kind of ingestion process, and how much automated is the entire ingestion process, type of access to the dataset, overtime, period, associated parameters, date of last update, date of triples creation, status of the ingestion process, etc.

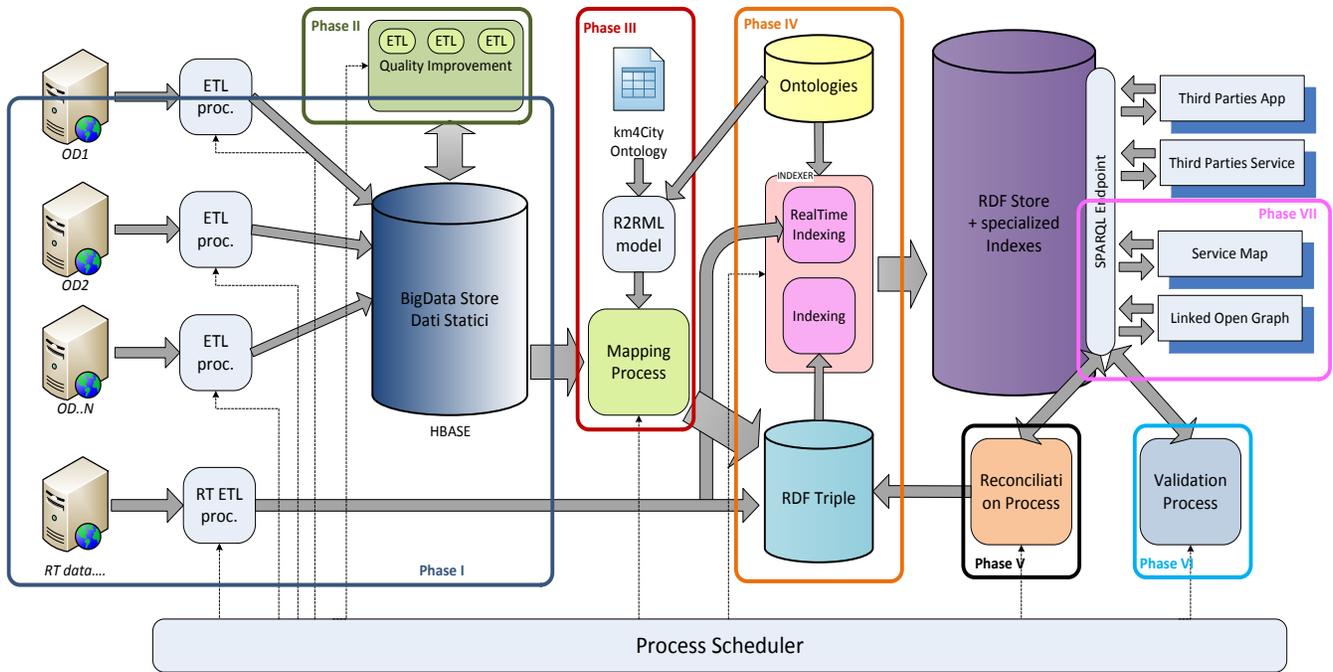

Figure 4 – Big Data Processing Architecture Overview

## IV. DATA ENGINEERING ARCHITECTURE

In this section, the description of the data engineering architecture is presented (see Figure 4). The whole ingestion and quality improvement process can be regarded as divided into the following phases of: Data Ingestions (Phase I), Quality Improvement (Phase II), Mapping (Phase III), Indexing (Phase IV), Reconciliation (Phase V) to make the model semantic interoperable, Verification and Validation (Phase VI) and Access/exploitation from services (Phase VII). The whole phases of the ingestion processes are managed by a Process Scheduler that allocates processes on a parallel and distributed architecture composed by several servers. To allow the regular update of ingested data the scheduler regularly retrieves data and check for updates. The ingested data are transcoded, improved and then mapped in the km4City Ontology. After that, they are made available to applications on an RDF Store (OWLIM-SE) using a SPARQL Endpoint. Applications can use the geo-referenced data to provide advanced services to the city citizens. Present applications, for knowledge base browsing via Linked Open Data (http://log.disit.org) and the Service Map (Http://servicemap.disit.org), described in the following section, can be used as explicative and services for providing examples and SPARQL queries to enable the construction of third party applications on browsers and mobiles. ServiceMap and LOG are at the same time demonstrators and developing tools for who want to create new services through the project's APIs released; in fact, thanks to these tools, developers can better understand how the data are organized and how they can be used to produce useful service for citizens.

### A. Data Ingestion and Quality Improvement

For the data ingestion, the typical problems are related to the management of the several formats and of the various data sets that may find allocation on different segments and areas of the km4City Ontology, and that may be not semantically interoperable. The solution allows ingesting and harvesting a wide range of public and private data, coming as static, semi-static and real time data as mentioned in the previous sections. For the case of Florence area, we are addressing about 180 different sources of the 650 available. **Static and semi-static data** include POIs, geo-referenced services, maps, accidents statistic, and many statistics in general, etc. This information is typically accessible as public files in several formats, such as: SHP, KMZ, CVS, ZIP, XML, etc.

Each Open Data ingestion process retrieves information and produce records in a noSQL Hbase for big data [9], logging all the information acquired to trace back and versioning the data ingestion. Data are then completed; other columns are updated dynamically with other process steps, and finally data obtained are placed on an HBase table. Each open data has its own ingestion process consisting of a scheduled ETL transformation (Extract, Transform, and Load) on a parallel and distributed architecture.

**Real time data** includes data coming from sensors (e.g., parking, weather conditions, pollution measures, position of busses, etc.) that are typically acquired from Web Services also by using a scheduled and specific ETL transformation process. In most cases, the real-time data are directly pushed in the mapping process to feed the temporary SQL store. They are typically streamed into the traditional SQL store and then converted into triples in the RDF final store.

In almost all cases, each single data set is ingested by means of a different ETL process defined by using Pentaho Kettle formalism [10] because, among the several existing solutions, this formalism is quite diffused and easier to understand, and it was already used by Information Systems Directorate of Florence. When the Kettle language presented limitation, external processes in Java have been adopted.

Once stored on HBase, a process of **Quality Improvement** is applied on these data, which aims to improve the quality of the data before they will be transformed in triples, however allowing keeping track of their original shape. This Quality Improvement phase is realized by means of a set of ETL transformations, created as a result of an in-depth analysis on the most frequent errors contained in each ingested dataset and thus mainly for specific data types as dates, time, locality, addresses, URL, email, telephones and fax, CAP code, ateco codes, etc.

*B. Data Mapping and Indexing*

The **Mapping** (Phase III) deals with the transport of information, previously saved and polished into HBase database, into an RDF datastore, in our case managed by Owlim-SE [11]. The first part of this procedure retrieves information from HBase to put them on a temporary MySQL database (required to use the Data Integration tool chosen), while in the second part data are translated into triples. Transformation is needed to map the traditional structured into RDF triples, based on information contained in a well-defined km4City ontology and all ontologies reused (dcterms, foaf, schema.org, etc.). This process may be performed by ad-hoc programs that have to take into account the mapping from linear model to RDF structures. This two steps process allowed us to test and validate several different solutions for mapping traditional information into RDF triples and ontology. The ontological model has been several times updated and thus the full RDF storage has been regenerated from scratch reloading the definition (all the other vocabularies, selecting the testing several different solutions) and the instance triples according to the new model under test. Once the model has been generated, triples can be automatically inserted.

The first essential step is to specify semantic types of the data set, i.e., it is necessary to establish the relationship between the columns of the SQL tables and properties of ontology classes. The second step consists in defining the Object Properties among the classes, or the relationships between the classes of the km4City ontology. When dataset has 2 columns that have the same semantic type but which correspond to different entities, thus multiple instances of the same class have to be defined, associate each column to the correct one.

The process responsible to perform the mapping transformation, passing from Hbase to SQL database has been produced as a corresponding ETL Kettle associated with each specific ingestion procedure for each data set. The second phase, of performing the mapping from SQL to RDF, has been realized by using a mapping model: Karma Data Integration tool [12], which generates a R2RML model, representing the mapping for transport from MySQL to RDF and then it is uploaded in a OWLIM-SE RDF Store instance [11]. Karma initialization phase involves loading the primary reference ontology and connecting dataset containing the data to be mapped. This process allowed the production of the knowledge base that may present a large set of problems due to inconsistencies and incompleteness that may be due to lack of relationships among different data sets, etc. These problems may lead to the impossibility of making deductions and reasoning on the knowledge base, and thus on reducing the effectiveness of the model constructed. These problems have to be solved by using a reconciliation process (Phase V) via specific tools and the support of human supervisors.

Moreover, once the mapping is performed a large set of triples are available coming from: ontologies, static data, real time historical data, quality improvements, real time data. These triples have to be loaded into the RDF stores and specific indexes have to be built to make the store suitable for real time queries. Thus, Phase IV of **indexing** is performed. Once the RDF store is ready, specific algorithms for detecting entities to be reconciled can be executed exploiting the SPARQL entry point as described in the following.

*C. Exploiting and Exploriing Smart City Data*

The km4City Ontology presented in this paper is a strong generalization of a large set of data modeling problems. The integration of the several data sets coming from different sources into a semantic interoperable knowledge base is a solution to exploit this information for smart city purpose. To this end, the activities of data quality improvements can be performed in Phase II after the ingestion, and/or may be after the triple generation and indexing to discover problems of reconciliation and to solve them.

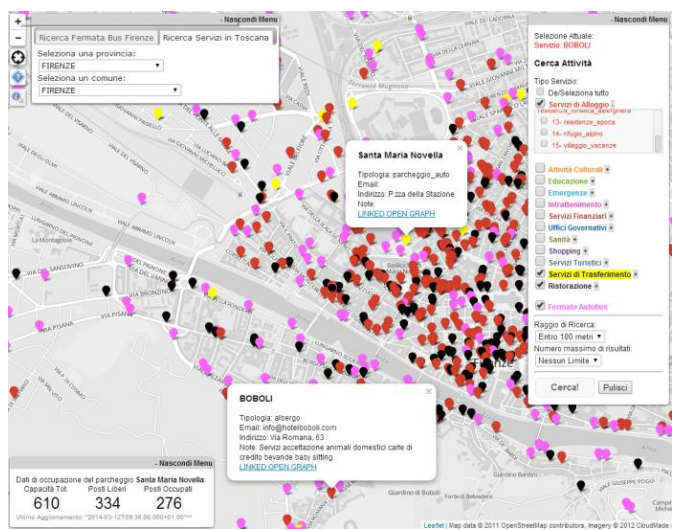

Figure 5 - Service Map (http://servicemap.disit.org)

The system has been used to ingest the data coming from the Municipality of Florence, the Tuscany Region and MIIC. Considering only files related to the daily weather forecast of all the available municipalities, we have 286 files updated twice a day, each of which, containing also 16 lines of weather prediction for the week, we obtain an increase of approximately 270,000 HBase lines per month that, in terms of triples, corresponds to a monthly increase of about 4 million triples.

Moreover, in order to explore the data being ingested and their relationships a tool for data visualization and exploration was used, that allows exploring the semantic graph of the relations among the entities, this Linked Open Graph [17] is available for applications developers to explore and understand better the data available in the ontology.

A second tool called ServiceMap to perform geographic queries has been deployed. It can be used to perform geographic queries and getting the POIs close to a bus station, to a street number, to a given point on the map, etc. (see Figure 5). The service map, for example, allows to (i) get bus stops and from them to access the status line of the bus, providing the waiting time to the next bus, (ii) finding parking and getting in real time the number of empty places, etc. From each "pin" in the map, it is possible to pass from the entity identified to its model in terms of relationships on the LOG.disit.org graph [17].

## V. VERIFICATION, VALIDATION AND RECONCILIATION

To connect services to the Street Guide in the repository a reconciliation phase in more steps, has been required, because the notation used by the Tuscany region in some Open Data within the Street Guide, does not always coincide with those used inside Open Data relating to different points of interest. In substance, different public administrations are publishing Open Data that are not semantically interoperable.

Typical problems can be related to: (i) low quality of data, (ii) lack of data that are supposed to arrive in real time, (iii) changes in the data model of the data set, (iv) changes and updates into the data sets (this problem could generate a change into the ontological model and thus the human intervention is activated for model review), etc. To this end, periodic verification and validation processes is needed to be performed by defining a set of SPARQL queries on the knowledge base with the aim of detecting inconsistencies and incompleteness, and verifying the correct status of the model. These periodically executed queries perform a regression testing every time a new update of data process ingestion is performed, and when real time data arrive into the final RDF store. The validation process may lead to identify problems that may be limited to the instances of classes. To this end, the fourth information associated with each triple allows to identify the problems and the data set processes to be revised.

Therefore, an iterative workflow process was defined. During validation there were cases like the Weather forecast where no connection among the data were present due to different encoding of the name of the municipality, for this reason to support the reconciliation process a table containing the ISTAT code of each municipality was created, and each time new weather data are available, they are automatically completed with the correct ISTAT code, thus supporting the search for the instance of the PA class to which connect the weather forecasts.

A relevant process of data improvement for semantic interoperability is related to the application of reconciliations among the entities associated with locations as streets, civic numbers and localities. On this regard, there are different types of inconsistencies within the various integrated dataset, such as:

- typos;
- missing street number, or replacement with "0" or "SNC" (Italian acronym that means without civic number);
- Municipalities with no official name (e.g. Vicchio/Vicchio del Mugello);
- street names with uncommon characters ( -, /, ° ? , Ang., ,);
- street numbers with strange characters ( -, /, ° ?, Ang. ,(, );
- road name with words in a different order from the usual ( e.g. Via Petrarca Francesco, exchange of name and surname);
- number wrongly written (e.g. 34/AB, 403D, 36INT.1);
- red street numbers (in some cities, street numbers may have a color. So that a street may have 4/Black and 4/Red, red is the numbering system for shops);Roman numerals in the road name (e.g., via Papa Giovanni XXIII).

As a summary, the whole knowledge base initially created was consisting of more than 81 Million triples, with a growth of 4 million triples per month. A part of them can be discharged when statistical values are estimated and punctual value discharged. For the validation, a total amount of services/points of interest inserted into the repository has been of 36777 instances. Among these, 13185 have been reconciled at street number-level, while the number of elements reconciled at street-level has been 21207, all of them for the services. There are also 149 services associated to a coordinate pair, for which reconciliation did not return results, yet for the lack of references into the knowledge base (some streets and civic numbers are still missing or incomplete).

Thanks to the created ontology, is possible to perform services reconciliation at street number level, i.e. connecting an instance of class *Service* to an external access that uniquely identifies a street number on a road, or only at street-level, with less precision (lack that can be compensated thanks to geolocation of the service).

In the collected data sets, an average of about the 15% are automatically connected entities since they refer to perfectly consistent locations (i.e., perfect match in terms of location, street and civic number) in the MIIC with respect to the description reported in the service data set. In the total of location entities ingested, 5,75 % of locations are wrong and not reconciliable due to (i) the presence of wrong values for streets and/or locations, and (ii) the lack of a consistent reference location into the MIIC geographical model.

The reconciliation process can be performed with the aim of finding elements that identify the same entity while presenting different URIs. Thus the identified reconciliations are solved creating an *owl:sameAs* triple to the selected location toponym. Reconciliation detection can be performed by using (i) a set of specific SPARQL queries, (ii) program tools for RDF link discovering. To this end, declarative languages for link discovering such as SILK [14] and LIMES [13] have been proposed. As the production of SPARQL queries, the programming of the link discovering algorithms also implies the knowledge of the ontological structure of the RDF stores to be compared/linked.

### A. SPARQL Reconciliation

The methodology used for SPARQL reconciliation consists of trying to connect each service at street number-level, and then, perform the reconciliation at street-level. The first reconciliation step performed consists of an exact search of the street name associated with each service integrated. For example, to reconcile the service located at "VIA DELLA VIGNA NUOVA 40/R-42/R, FIRENZE", a SPARQL query is necessary, to search for all elements of *Road* class connected to the municipality of *"FIRENZE"* (via the ObjectProperty *inMunicipalityOf*), which have a name that exactly corresponds to "VIA DELLA VIGNA NUOVA" (checking both fields: official name, alternative name). The query results has to be filtered again, imposing that an instance of *StreetNumber* class exists and it corresponds to civic number "40" or "42", with the R class code Red. A very frequent problem for exact search, is the existence of multiple ways to express toponym qualifiers, that is dug (e.g. Piazza and P.zza) or parts of the proper name of the street (such as Santa, or S. or S or S.ta): thanks to support tables, inside which the possible change of notation for each individual case identified are inserted, a second reconciliation step was performed, based on exact search of the street name, which has allowed to increase the number of reconciled services at street number-level. The third reconciliation step is based on the research of the last word inside the field *v:Street-Address* of each instance of the *Service* class, because, statistically, for a high percentage of street names, this word is the key to uniquely identify a match.

The above mentioned three steps have been also carried out without taking into account the street number, and so in order to obtain a reconciliation at street-level of each individual service. An additional, phase of *manual correction* has been also performed by manually (i) searching services and incongruent locations via web search service as Google, (ii) cleaning address and street number fields, (iii) accepting and performing association match of non-identified matches, taking into account the list of probable candidates suggested by the query results.

*B.   Link discovering Based Reconciliation and comparison*

Link discovering based reconciliation consists in writing specific SILK algorithms for link discovering. They allow to discover links by writing specific algorithms grounded on distances and similarity metrics between patterns and relationships mainly based on string matching and distance measures (*Euclidean, weighted models, tree distances, patterns distance, string match, taxonomical, Jaro, Jaro-Winkler, Leveisthein, Dice, Jaccard, etc.*) [14].

In this case, a number of link discovering algorithms have been developed and assessed. Among them, the better ranked were based on comparing, at the same time, the location and the street. Firstly searching for the perfect match on location name and accepting uncertainty on street number from 0 up to 5 characters, for example. Both criteria have been aggregated considering their weight almost identical.

*C.   Reconciliation Comparison*

The obtained results are reported in Table 1. The table reports the results assessed in terms of precision, recall and F1score (the F1 score is also called the F-measure, and it is defined as Harmonic mean of Recall and Precision) [15], in identifying the correct entities to be reconciled. The first two lines refer to the SPARQL approach with and without manual intervention as described in Section V.A. The manual intervention has strongly improved the recall. On the other hand, the SPARQL approach is very time intensive for the programmers since a set of specific queries have to be produced for each data set to be reconciled. The second part of Table I reported the results related to different implementations of link discovering SILK based solutions, by using different string distances (i.e., Leveisthein, Dice, and Jaccard), with the above mentioned values for their parameters. Other distance models have been also used without obtaining significant results. The last Link discovering solution has been coded by using an additional knowledge about all the specific strings coding problems reported in Section V.

Table I – Reconciliation Comparison

| Method | Precision | Recall | F1 |
|---|---|---|---|
| SPARQL –based reconciliation | 1,00 | 0,69 | 0,820 |
| SPARQL -based reconciliation + manual action | 0,985 | 0,722 | 0,833 |
| Link discovering - Leveisthein | 0,927 | 0,508 | 0,656 |
| Link discovering - Dice | 0,968 | 0,674 | 0,794 |
| Link discovering - Jaccard | 1,000 | 0,472 | 0,642 |
| Link discovering - Knowledge base + Leveisthein | 0,925 | 0,714 | 0,806 |

*D.   Consideration on database size*

Table II shows the size, in terms of triples, of each macroclass area. As it can be seen from the numbers, the street-guide data corresponds to about half of all triples contained in the repository; furthermore, the only two RealTime macroclasses, i.e. Sensors and Temporal, are another important piece of information contained on triplestore.

Table II – Database numbers vs km43city macroareas and data kind

| Macroclass | Static Triples | Real Time Triples | Reconciliation Triples | Total Row |
|---|---|---|---|---|
| Administration | 2431 | 0 | 0 | 2431 |
| Local Public Transport | 644400 | 0 | 2385 | 646785 |
| Metadata | 416 | 0 | 0 | 416 |
| Point of Interest | 471657 | 0 | 34392 | 506049 |
| Sensors (busses and sensors) | 0 | 44111078 | 0 | 44111078 |
| Street-guide | 68985026 | 0 | 0 | 68985026 |
| Temporal | 0 | 8715105 | 0 | 8715105 |
| Total | 70103930 | 52826183 | 36777 | 122966890 |

In fact, RealTime data are closely related to the number of enabled services by each municipality or by the Tuscany region, considering that currently services architecture built AVM detected, corresponding to a small portion of all the bus lines (about a tenth of the total number of monitored lines). The number of real time triples reported are related to an year of processing. When all the bus lines will be available the real time triples would become 10 times bigger, dominating the size of the RDF store. On the other hands, for all the present services the full set of data about the bus position in each time instant of the day for several past months would not be needed. And thus, these details about bus data can be dropped to be substituted monthly with statistical values for the day, week and month for the delay. Therefore, also a strategy for the triplestore has been defined planning a reindexing every 4 months. Maintaining a couple of months of historical window on the triplestore. This allows us to keep under control the performance degradation, remaining always in the range of 90-120 million of triples. Anyway, the RealTime data, not reloaded into the last re-indexing triplestore, remain stored in terms of triples files for each generation in the RDF data store of Phase IV servers.

## VI. CONCLUSIONS

In this paper, a system for the ingestion of public and private data for smart city with related aspects as road graph, services available on the roads, traffic sensors etc., has been proposed. The system includes both open data from public administration and private data coming from transport systems integrated mangers, thus addressing and providing real time data of transport system, i.e., the busses, parking, traffic flows, etc. The system allows managing large volumes of data coming from a variety of sources considering both static and dynamic data. This data is then mapped to the km4City Ontology and stored into an RDF-Store where this data are available for applications via SPARQL queries to provide new services to the users (accessible KM4City document http://www.disit.org/5606 , in terms of schema http://www.disit.org/km4city/schema/ and in terms of OWL http://www.disit.org/6506 ). The derived ontology has been obtained by means of an incremental process performed analyzing, integrating and validating each added data set. Thus the resulting ontology is a strong generalization of a large set of data modeling problems.

In addition, a thorough verification and validation process performed allowed us to identify the set of triples to: (i) improve and enrich the model, and (ii) perform the corrections. Thus improving and enabling the deductive capabilities of the final model. Finally, the proposed system also provides a visualization and exploration tool to explore the data available in the RDF-Store. As a conclusion, the performed assessment and comparison has produced a clear results demonstrating that the best quality of results are obtained by using the approach based on SPARQL queries plus some manual actions. Also the simple usage of SPARQL queries resulted to be better ranked with respect to the SILK based link discovering. On the other hand, the writing of link discovering algorithms resulted to be much simpler and faster that performing a set of specific SPARQL queries. The next step will be to identify famous names, points of interest, locality names that can be linked to other data set as DBpedia[2] or GeoNames[3] according to a Linked Open Data model. This process can be performed with a simple NLP algorithm [6], [8].

---

[2] http://dbpedia.org/
[3] http://www.geonames.org/

ACKNOWLEDGMENT

Sincere thanks to the public administrations that provided the huge data collected and to the Ministry to provide the funding for Sii-Mobility Smart City Project, a warm thanks to Lapo Bicchielli, Giovanni Ortolani, and Francesco Tuveri.